\documentclass[prd,twocolumn,showpacs,floatfix,amsmath,nofootinbib,amssymb,floatfix]{revtex4}

\usepackage{epsfig}
\usepackage{longtable}
\usepackage{amsmath}
\usepackage{mathrsfs}
\usepackage{amssymb}
\usepackage{graphicx,color,dcolumn,booktabs,bm}
\usepackage{longtable,lscape,booktabs}
\usepackage{multirow,booktabs}
\usepackage{txfonts}
\usepackage{overpic}
\usepackage{siunitx}
\usepackage{indentfirst}
\usepackage{feynmf}   
\usepackage{slashed}  
\usepackage{cases}
\usepackage{color}
\usepackage{multirow}
\usepackage{epstopdf}
\usepackage{xcolor}
\usepackage{soul}
\usepackage{graphicx,color,dcolumn,booktabs,bm}
\usepackage{tabularx}
\usepackage[colorlinks,
            citecolor=blue,
            anchorcolor=red,
            menucolor=red,
            linkcolor=red,
            filecolor=red,
            runcolor=red,
            urlcolor=blue,
            frenchlinks=false]{hyperref}

\usepackage{soul}

\newcommand{\apjs}{ApJ} 

\begin{document}

\title{Neutrino and Cascade Gamma-Ray Emission from Magnetized Turbulent Coronae in Seyfert Galaxies}
\author{Xing-Jian Wang $^{1,2}$}
 \author{Jing-Fu Hu$^{1}$}\email{hujingfu@qhnu.edu.cn}
 \author{Hao-Ning He$^{2}$}\email{hnhe@pmo.ac.cn}
 \author{Cheng-Qun Pang$^{3,4}$}
\affiliation{
$^1$College of Physics and Electronic Information Engineering, Qinghai Normal University, Xining 810000, China\\
$^2$Key Laboratory of Dark Matter and Space Astronomy, Purple Mountain Observatory, Chinese
 Academy of Sciences, Nanjing 210023, China\\
 $^3$School of Physics and Optoelectronic Engineering, Ludong University, Yantai 264000, China\\
$^4$Lanzhou Center for Theoretical Physics, Key Laboratory of Theoretical Physics of Gansu Province, Lanzhou University, Lanzhou, Gansu 730000, China}
\begin{abstract}
Recent neutrino observations from the IceCube Collaboration suggest that Seyfert galaxies are promising candidate sources of neutrinos. Within the standard disk–corona model, we assume that protons are accelerated by a non-resonant acceleration mechanism driven by magnetized turbulence in the corona. These accelerated protons interact with ambient radiation or matter, producing high-energy neutrinos and gamma rays. In this scenario, gamma rays are largely absorbed within the corona. The neutrino luminosity depends primarily on the properties of the corona (such as the X-ray luminosity and radius) and the spectral energy distribution of the target photons. This study demonstrates the relation between the neutrino luminosity and the X-ray luminosity, and further discusses the contribution of cascade gamma rays to coronal radiation. Notably, MeV gamma rays can effectively escape the source, together with neutrinos, and serve as key observational probes for testing this model. Future MeV gamma-ray telescopes, such as AMEGO-X and e-ASTROGAM, are expected to detect such gamma-ray signatures, providing a critical multi-messenger test of the hadronic corona model.
\end{abstract}
\maketitle

\section{Introduction}

Seyfert galaxies have long served as pivotal natural laboratories for unraveling the links between accretion dynamics, particle acceleration, and high-energy radiation in the Universe. The recent detection by IceCube of excesses of high-energy neutrinos from Seyfert galaxies NGC 1068, NGC 4151 and CGCG 420-015~\cite{IceCube:2022der, IceCube:2024dou, Neronov:2023aks, Sommani:2024sbp} has provided evidence that Seyfert galaxies are promising neutrino source candidates.
Neutrino production typically requires interactions of high-energy protons with ambient radiation or matter.
However, it remains unclear where and how these protons are accelerated in active galactic nuclei (AGN).
Several acceleration mechanisms have been proposed, such as accelerations in elativistic jets~\cite{Murase:2014foa, Rodrigues:2017fmu},  stochastic acceleration by turbulence or magnetic reconnection in the corona~\cite{Inoue_2019, Kheirandish:2021wkm}, shock acceleration from infalling matter or failed winds~\cite{Inoue:2019yfs}, and direct acceleration in black hole magnetospheres~\cite{Neronov:2009zz}.

The disk-corona model offers a compelling framework to interpret these neutrino signals~\cite{Murase:2019vdl, Fiorillo:2024akm, Das:2024vug}. 
In this model, Comptonized X-rays from the corona provide target photons for photomeson production~\cite{Galeev:1979td, 1976ApJ...206..910K}.
Early work by \citet{Murase:2019vdl} explored neutrino and cascade photon emission assuming resonant stochastic proton acceleration in the corona. 
However, recent particle-in-cell (PIC) simulations indicate that stochastic acceleration is predominantly non-resonant~\cite{Comisso:2019frj, Comisso:2024ymy}. 
Unlike resonant processes, this mechanism features an energy-independent acceleration timescale governed by the largest-scale turbulent structures.
Building on this, ~\citet{Fiorillo:2024akm} demonstrated that non-resonant acceleration in a highly turbulent corona can yield neutrino fluxes consistent with IceCube observations. 

A natural consequence of coronal models is a correlation between neutrino and X-ray luminosities, as both originate from the same underlying energy source~\cite{Neronov:2025cfc}.
While models based on resonant acceleration can relate neutrino emission to X-ray luminosity~\cite{saurenhaus2025constrainingcontributionseyfertgalaxies}, they tend to overpredict the diffuse TeV neutrino background if all sources resemble NGC 1068. 
In contrast, the specific model of~\citet{Fiorillo:2025ehn}, which implements non-resonant acceleration, successfully reproduces both the observed diffuse flux at several TeV and a scaling relation between neutrino and X-ray luminosities.

Gamma rays are expected to be produced at a flux level comparable to neutrinos. 
However, multi-messenger observations of NGC 1068 and NGC 4151 reveal a gamma-ray flux significantly suppressed relative to the neutrino flux~\cite{IceCube:2022der, IceCube:2019cia, Ajello:2023hkh, MAGIC:2019fvw, MAGIC:2025tlp}. 
This discrepancy can be explained if the gamma-ray emission site, such as the corona, is opaque to high-energy photons~\cite{Murase:2022dog}.
Within the corona, absorbed high-energy gamma rays can initiate electromagnetic cascades, leading to MeV-range secondary emission with a luminosity comparable to that of neutrinos~\cite{Neronov:2023aks, Kun:2024meq}.
Consequently, MeV emission can serve as a viable probe for identifying neutrino source candidates.
Future sensitive MeV telescopes such as the All-sky Medium Energy Gamma-ray Observatory eXplorer (AMEGO-X)~\cite{Caputo:2022xpx} and e-ASTROGAM~\cite{e-ASTROGAM:2016bph} will allow for a direct quantification of the contribution of Seyfert galaxies to the astrophysical neutrino flux.

In this paper, we investigate the correlation between MeV cascade emission and neutrino emission for the population of Seyfert galaxies, providing predictions for future MeV gamma-ray and neutrino observations.
Adopting the framework of coronal properties and the non-resonant acceleration mechanisms described in~\cite{Fiorillo:2024akm, Fiorillo:2025ehn}, we explore the relationship between the neutrino luminosity and the X-ray luminosity, with a particular focus on the implications of the electromagnetic cascade.
The structure of the paper is as follows. In Section~\ref{sec: model}, we develop the disk-corona model using a set of theoretical and empirical relations. In Section~\ref{sec: non-thermal proton}, we describe non-thermal protons accelerated via magnetized turbulence in the corona. We show our results on the properties of the Seyfert galaxy population in Section~\ref{sec: neutrino and gamma} and the summary in Section~\ref{sec: summary}. Throughout this paper, we use the notation $Q_{x} = Q / 10^{x}$ in CGS units.

\section{Disk-corona model}\label{sec: model}

In this model, the accretion disk is modeled as a standard optically thick, geometrically thin Shakura-Sunyaev disk. Its radius spans a range from $3R_s$ to several hundred $R_s$, where $R_s = 2GM_{\rm BH}/c^2$ is the Schwarzschild radius, $M_{\rm BH}$ represents the mass of the central supermassive black hole (SMBH), $G$ denotes the gravitational constant, and $c$ is the speed of light. 
The resulting spectrum is characterized as a multi-temperature blackbody spectrum. The local temperature $T(r)$ at radius $r$ is given by~\cite{Ghisellini:2009wa}
\begin{equation}
    T^{4}\left(r \right) = \frac{3 R_{s} L_{\rm d}}{16 \pi \eta_{\rm rad} \sigma_{\rm SB} r^3} \left[ 1 - \left(\frac{3 R_{s}}{r}\right)^{1/2} \right],
\end{equation}
where $\eta_{\rm rad} = 0.1$~\cite{Lasota:2015cga} denotes the radiative efficiency, $L_{\rm d} \simeq 0.5 L_{\rm bol}$~\cite{Murase:2019vdl} represents the disk luminosity with $L_{\rm bol}$ being the bolometric luminosity, and $\sigma_{\rm SB}$ is the Stefan-Boltzmann constant. 
Integrating over the disk within radius R, we obtain the disk's contribution to the target photon field as described by
\begin{equation}
    \nu F_{\nu} = \frac{8 \pi^2 h \nu^3}{c^2} \int_{3R_s}^{R} \frac{r}{\exp \left(\frac{\varepsilon}{k_{B} T(r)}\right) - 1} \,dr,
\end{equation}
which exhibits a characteristic 'big blue bump' in the optical-UV range.

The corona is approximated as a spherical, magnetized plasma region surrounding the SMBH with a radius $R = \mathcal{R} R_s$, where $\mathcal{R}$ is the dimensionless normalized radius. 
We assume the electron density is governed by the Compton opacity, i.e.,
$n_e\approx \tau_{T} / (\sigma_{T} R)$, where $\tau_{T}$ is the Compton optical depth and $\sigma_T$ denotes the Thomson cross section. This assumption links the coronal compactness with its X-ray emission. 
The magnetic field strength is estimated by assuming equipartition with the gas pressure, 
characterized by the plasma parameter $\beta$, i.e.,
\begin{equation}
    B = \sqrt{\frac{8 \pi n_p k_B T_p}{\beta}},
\end{equation}
where the proton density is comparable to the electron density, i.e.,  $n_e \approx n_p$~\cite{Fiorillo:2024akm},
and the proton temperature is assumed to be at the virial temperature, $T_p = m_{p} c^{2} / (6 \mathcal{R} k_B)$~\cite{Murase:2019vdl}.

The X-ray emission from the corona is produced by the Comptonization of disk photons. Its spectrum can be modeled by a power law with an exponential cutoff,
\begin{equation}
    \frac{dn_{\gamma}}{d\varepsilon_{\gamma}} \propto \varepsilon_{\gamma}^{- \Gamma_{X}} \exp \left(- \frac{\varepsilon_{\gamma}}{\varepsilon_{X, \rm{cut}}} \right).
\end{equation}
The spectral index $\Gamma_{X}$ and the cutoff energy $\varepsilon_{X,\mathrm{cut}}$ are empirically correlated with the Eddington ratio $\lambda_{\rm Edd}$. Thus we have $\Gamma_{X} \approx 0.167 \times \log(\lambda_{\rm Edd}) + 2.0$~\cite{Trakhtenbrot:2017xiz} 
and $\varepsilon_{X,\mathrm{cut}} \approx \bigl[-74 \log(\lambda_{\rm Edd}) + 150\bigr]$ keV~\cite{Ricci:2018eir} with $\lambda_{\rm Edd} \equiv L_{\rm bol} / L_{\rm Edd}$, where $L_{\rm Edd}$ denote the Eddington luminosity. 
The intrinsic X-ray luminosity $L_{X}$ in the $2-10$ keV band is related to the photon density in the corona following 
\begin{equation}
    \int_{2\,\rm{keV}}^{10\,\rm{keV}} \varepsilon_\gamma \frac {dn_\gamma}{d\varepsilon_\gamma} \, d\varepsilon_{\gamma} \equiv \frac{(1+\tau_T)L_X}{4\pi R^2c},
\end{equation}
where the Compton optical depth $\tau_{T}$ can be calculated via $\tau_{T} = 10^{(2.16 - \Gamma_{X}) / 1.06} (k_{B} T_{e} /\, \rm{keV})^{-0.3}$ with the electron temperature $T_e\approx \varepsilon_{X, \rm{cut}}/(2k_{B})$ in a slab geometry conora model~\cite{Ricci:2018eir}.
Furthermore, the bolometric luminosity $L_{\rm bol}$ can be calculated via the relationship between $L_{\rm bol}$ and $L_{X}$~\cite{Hopkins:2006fq},
\begin{equation}
    \frac{L_{\rm bol}}{L_{X}} = 10.83\left(\frac{L_{\rm bol}}{10^{10}L_{\odot}}\right)^{0.28} + 6.08\left(\frac{L_{\rm bol}}{10^{10}L_{\odot}}\right)^{-0.02},
\end{equation}
where $L_{\odot} \equiv 3.9 \times 10^{33}~{\rm erg}~{\rm s}^{-1}$.

\begin{figure}[t]
\begin{center}
\includegraphics[width=\linewidth]{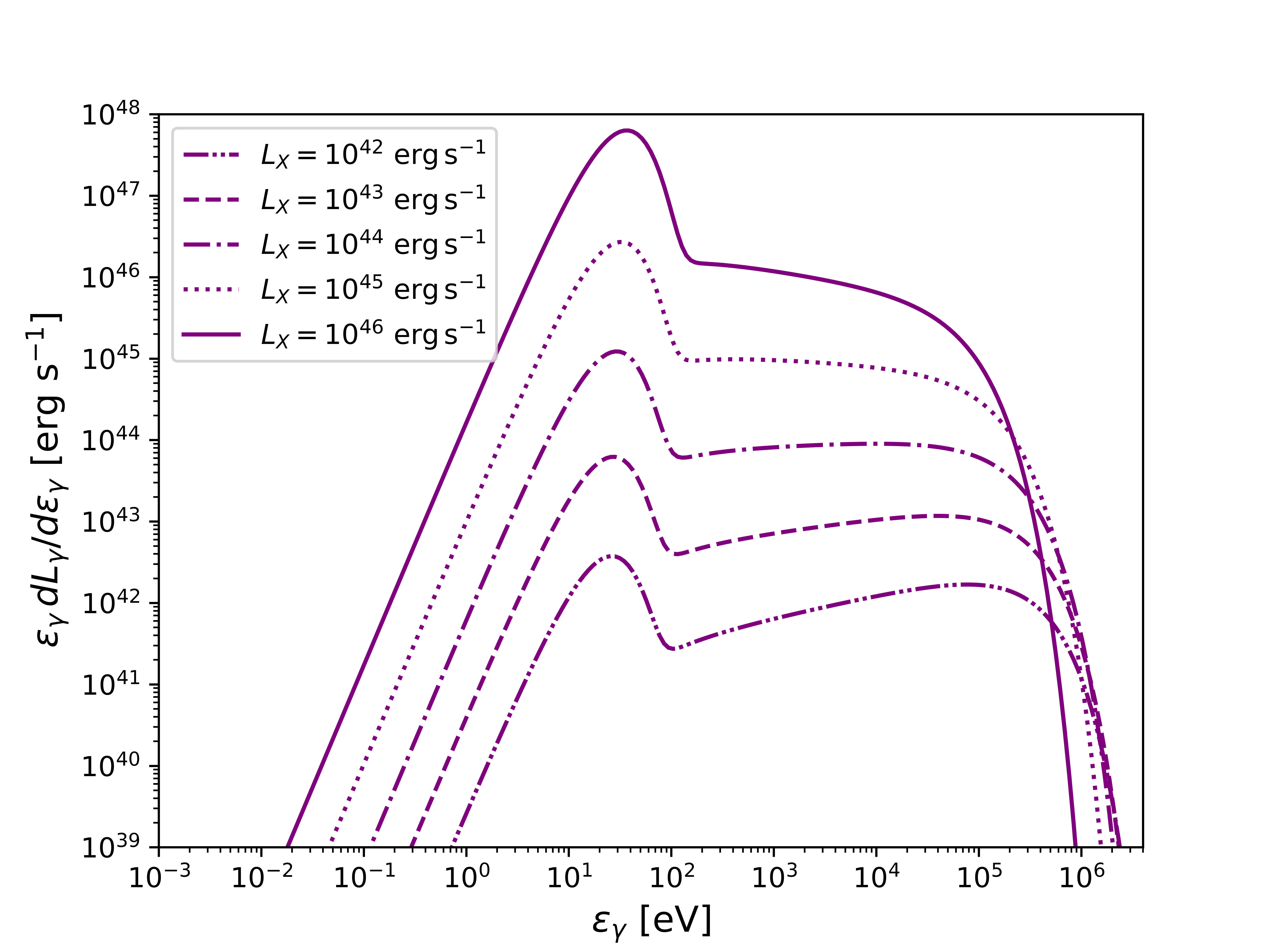}
\caption{
Spectral energy distributions of the target photon field under different X-ray luminosities correspond to the calculation results with parameter $\mathcal{R}=10$.
Curves for different X-ray luminosity values are labeled in the legend.
}
    
\label{fig:target}
    
\end{center}
\end{figure}

To close the mdoel, we estimate the central black hole mass $M_{\rm{BH}}$ using the empirical relation~\cite{Mayers:2018hau}
\begin{equation}
     M_{\rm{BH}} = 3 \times 10^7 M_{\odot} \left( \frac{L_X}{2 \times 10^{43}~{\rm erg}~{\rm s}^{-1}} \right)^{0.58},
\end{equation}
which connects $M_{\rm{BH}}$ to the X-ray luminosity. 

In Fig.~\ref{fig:target}, we present spectral energy distributions (SED) of the target photon field under the disk-corona model. The SED is characterized by two dominant components, a prominent \'big blue bump' peaking in the infrared–optical regime, originating from the multi-temperature blackbody emission of the accretion disk, and a non-thermal power-law component from the corona, which dominates the X-ray band and extends to energies of $\sim 10^3$ keV. This broadband photon field, particularly the dense X-ray component, serves as the primary target for proton energy losses and subsequent neutrino production via photomeson interactions.

\section{Non-thermal protons in the corona}\label{sec: non-thermal proton}

\begin{figure}[t]
\begin{center}
\includegraphics[width=\linewidth]{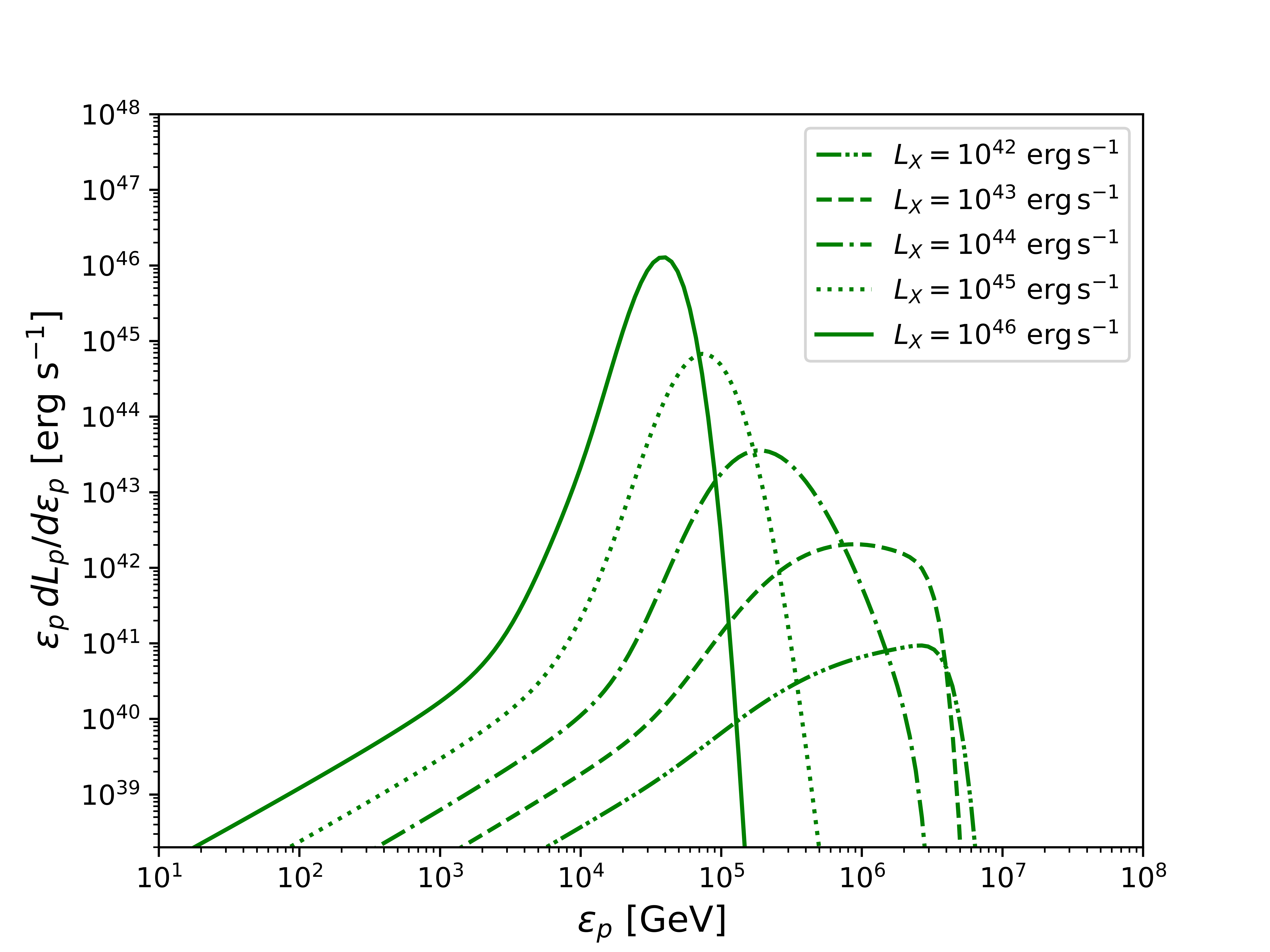}
\caption{Spectral energy distributions of non-thermal protons under different X-ray luminosities  correspond to the calculation results with parameter $\mathcal{R}=10$. Curves for different X-ray luminosity values are labeled in the legend.
}  
\label{fig:ProtonSpectrum}   
\end{center}
\end{figure}
\begin{figure*}[t]
\begin{minipage}{0.33\hsize}
\begin{center}
\includegraphics[width=\linewidth]{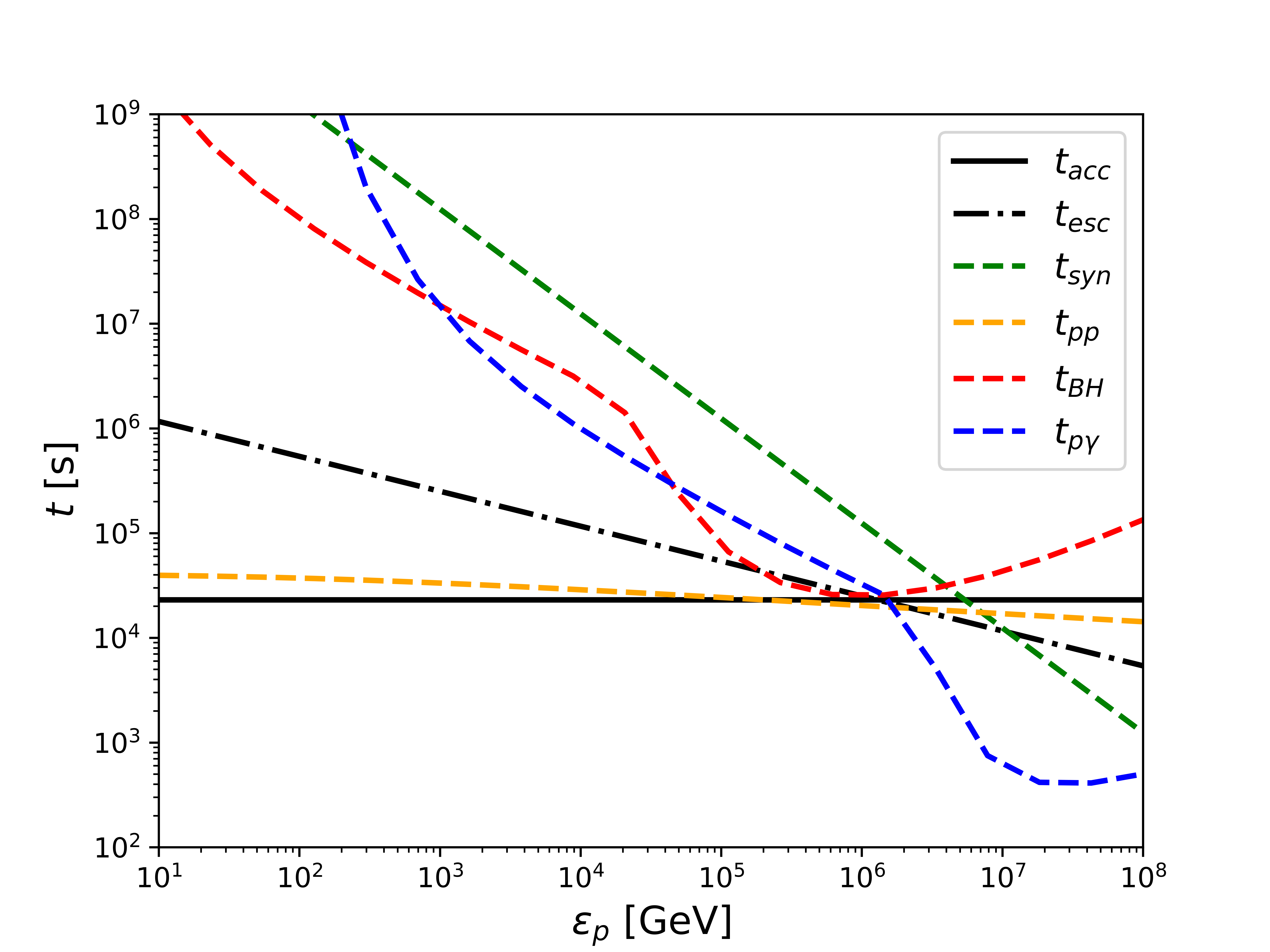}
\end{center}
\end{minipage}
\begin{minipage}{0.33\hsize}
\begin{center}
\includegraphics[width=\linewidth]{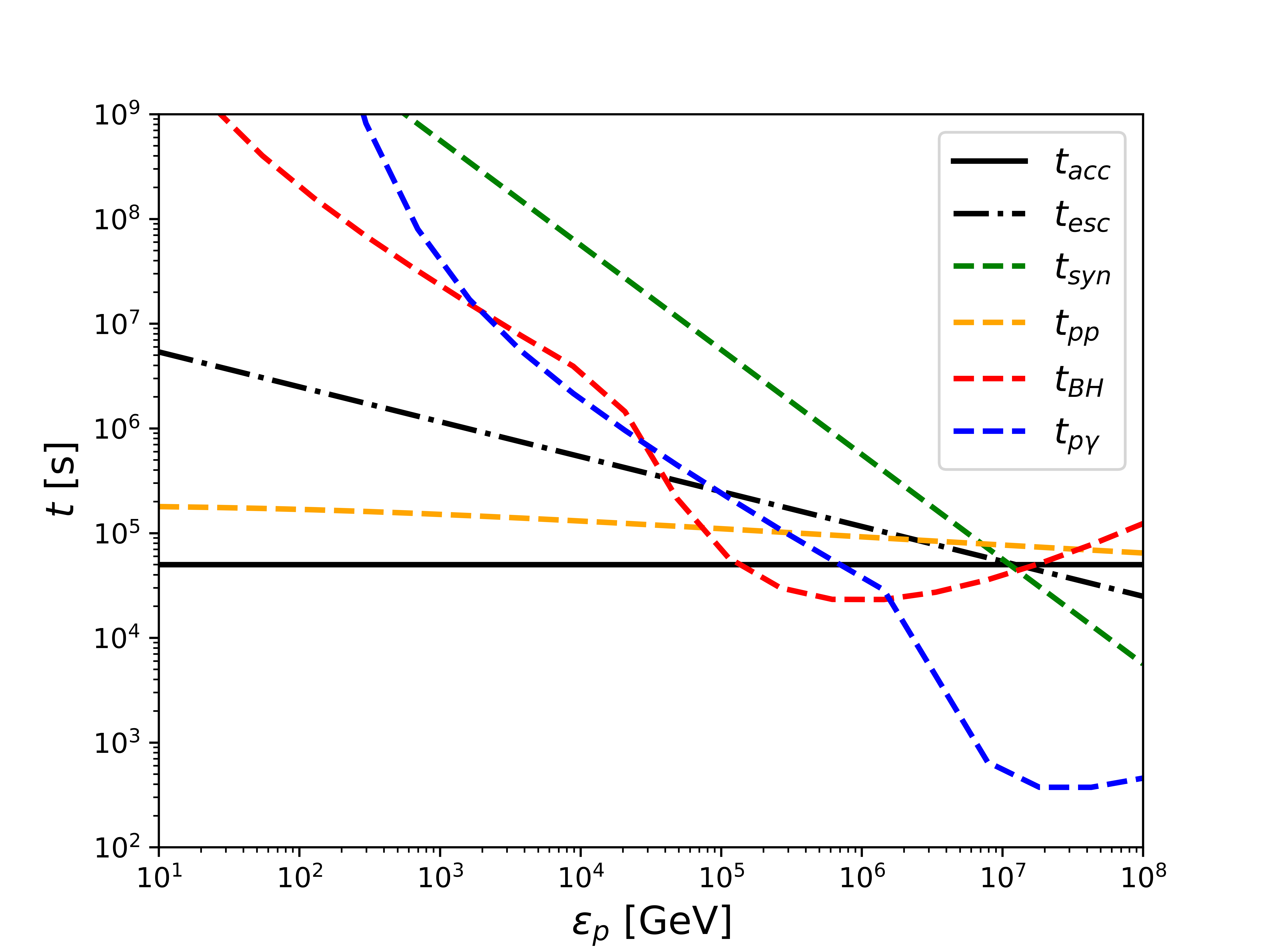}
\end{center}
\end{minipage}
\begin{minipage}{0.33\hsize}
\begin{center}
\includegraphics[width=\linewidth]{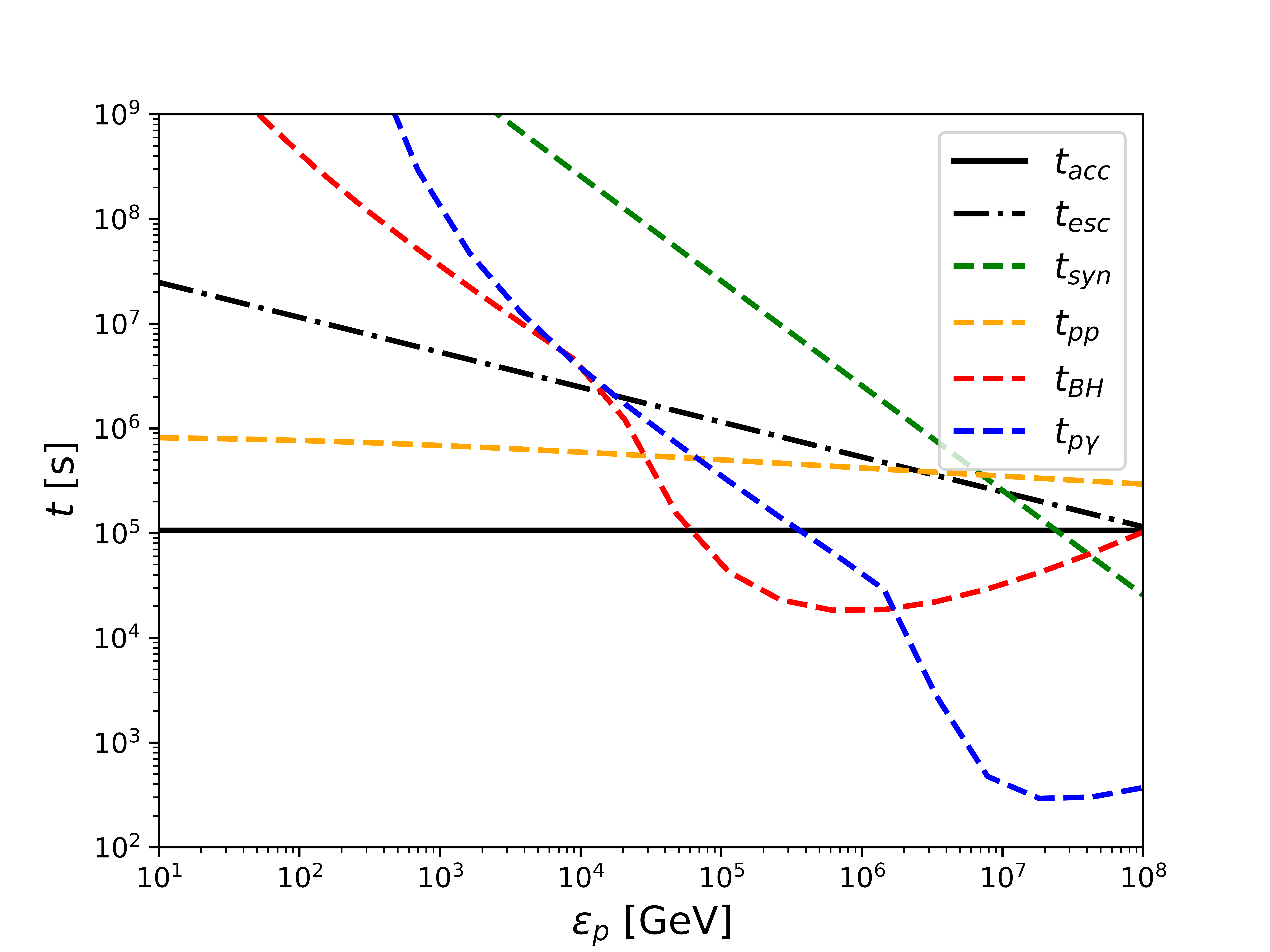}
\end{center}
\end{minipage}
\caption{
Timescales for different processes considered in this study, at $\mathcal{R}=10$ for different X-ray luminosity values of $L_{X}=10^{42}~{\rm erg}~{\rm s}^{-1}$ (left), $10^{43}~{\rm erg}~{\rm s}^{-1}$ (middle), and $10^{44}~{\rm erg}~{\rm s}^{-1}$ (right).
}
\label{fig:proton_times}
\end{figure*}
\setlength{\tabcolsep}{8pt}
\begin{table}[b]
\centering 
\caption{Parameters in our models. Units are $[\rm{eV}]$ for $\varepsilon_{\rm{inj}}$. \label{tab:quantities}}
\begin{tabular}{ccccc} 
\hline 
\multicolumn{5}{c}{Model-dependent parameters} \\ 
\hline 
$\beta$ & $\eta$ & $\mathcal{F}_{X}$ & $f_{\rm inj}$ & $\varepsilon_{\rm{inj}}$ \\ 
\hline 
\\
1 & 0.5 & 0.8 & $5\times 10^{-7}$ & $5 \times 10^{9}$ \\
\hline 
\end{tabular}
\end{table}
The production of high-energy neutrinos and gamma rays fundamentally requires a population of non-thermal protons. In this section, we describe the processes responsible for stochastically accelerating protons to high energies within the magnetized, turbulent corona, and the competing cooling mechanisms that shape their final energy distribution. The steady-state proton spectrum, $\mathcal{F}_p$, is the crucial link between the coronal properties discussed in Section \ref{sec: model} and the secondary particle emissions we calculate in Section \ref{sec: neutrino and gamma}.
The time evolution of the proton distribution function $\mathcal{F}_p$ in momentum space is governed by the Fokker-Planck equation
\cite{Kimura:2016fjx, Petrosian:2012ba},
\begin{equation}
\frac{\partial {\mathcal F}_p}{\partial t} = \frac{1}{\varepsilon_p^2}\frac{\partial}{\partial \varepsilon_p}\left(\varepsilon_p^2D_{\varepsilon_p}\frac{\partial {\mathcal F}_p}{\partial \varepsilon_p} + \frac{\varepsilon_p^3}{t_{p,\rm cool}}{\mathcal F}_p\right) -\frac{{\mathcal F}_p}{t_{\rm esc}}+\dot {\mathcal F}_{p,\rm inj}.
\label{eq:F_p}
\end{equation}
On the right-hand side, this equation accounts for the competition between several physical processes, the first term represents stochastic diffusion in energy space due to acceleration, with a diffusion coefficient $D_{\varepsilon_p}\equiv \varepsilon_p^2/t_{\rm acc}$, the second term is the advective loss due to continuous cooling processes, the third term describes particle escape from the acceleration region, and the final term is the injection of fresh particles into the accelerator.

Recent PIC simulations indicate that stochastic acceleration in magnetized turbulence is non-resonant \cite{Comisso:2019frj, Bresci:2022awc}. This marks a paradigm shift from the traditional resonant picture. Earlier theories linked the acceleration timescale to the proton gyroradius $r_{g}$ resulting in a strong energy dependence \cite{Kimura:2019yjo, Kakuwa:2015chr, Stawarz:2008sp}.
In the non-resonant regime, the acceleration is instead dominated by the largest-scale turbulent structures, resulting in an energy-independent acceleration timescale\cite{Comisso:2019frj, Fiorillo:2024akm}
\begin{equation}
    t_{\rm acc} \approx \frac{10}{\sigma_{\rm tur}} \frac{\ell}{c},
\end{equation}
where $\sigma_{\rm tur}$ is the turbulent magnetization parameter and $\ell= \eta R$ is the spatial scale of turbulence with $\eta < 1$. 
This key feature enables more efficient acceleration of protons to the high energies required for photomeson production, providing a natural explanation for the observed TeV–PeV neutrino signals.

To quantitatively apply this model, the turbulent magnetization $\sigma_{\rm tur}$ needs be determined. This is achieved by linking the observed X-ray output to the underlying magnetic energy dissipation. We define the fraction of turbulent magnetic power dissipated into X-rays as
$\mathcal{F}_{X} = L_{X, \rm{tot}} / L_{B}$~\cite{Fiorillo:2025ehn}, where $L_{X, \rm{tot}}$ is the total X-ray luminosity from 1 keV to the cutoff energy $\varepsilon_{X, \rm{cut}}$ and $L_{B}$ is the turbulent magnetic field energy dissipation rate, which is written as~\cite{Fiorillo:2024akm}
\begin{equation}
    L_{B} \approx \frac{2 \pi}{3} \frac{\eta_{\rm{rec}}}{\eta} \frac{c^3 \sigma^{3/2}_{\rm tur}}{(1 + \sigma)^{1/2}} (n_p m_p + n_e m_e) R^2,
\end{equation}
where $\eta_{\rm{rec}} \approx 0.1$ \cite{Cassak:2017enb, Comisso:2016ima} and 
$\sigma$ is the magnetization of the plasma, defined as
\begin{equation}
    \sigma \equiv \frac{B^2}{4 \pi (n_{p}m_{p} + n_{e}m_{e})c^2}.
\end{equation}
By assuming a characteristic value for $\mathcal{F}_{X}$, we use the measured $L_{X, \rm{tot}}$ to calculate $L_{B}$, and further solve $\sigma_{\rm tur}$ numerically, completing the parameter set for the acceleration model. This approach ensures that our turbulent acceleration scenario is energetically consistent with the observed coronal X-ray emission.
\begin{table}[b]
\centering 
\caption{Values of each physical quantity deduced in our model (for $\mathcal{R} = 10$) as a function of $L_{X}$. Units are [${\rm erg}~{\rm s}^{-1}$] for $L_X$, $L_{B}$ and $L_{p}$, [$M_{\odot}$] for $M$, and [$\rm cm^{-3}$] for $n_p$.}
\label{tab:physicalquantities} 
\begin{tabular}{c||cc|ccc}
\hline
 $\log L_X$ & $\log L_{B}$ & $\log L_{p}$ & $\log M$ & $\log n_p$ & $\sigma_{\rm tur}$ \\
 \hline
42.0 & 43.0 & 41.3 & 6.72 & 10.76 & 0.11 \\
43.0 & 43.9 & 42.7 & 7.30 & 10.10 & 0.20 \\
44.0 & 44.7 & 43.7 & 7.88 & 9.44 & 0.35 \\
45.0 & 45.6 & 44.8 & 8.46 & 8.80 & 0.64 \\
46.0 & 46.6 & 45.9 & 9.04 & 8.18 & 1.15 \\
\hline
\end{tabular}
\end{table}
\begin{figure}[b]
\begin{center}
\includegraphics[width=\linewidth]{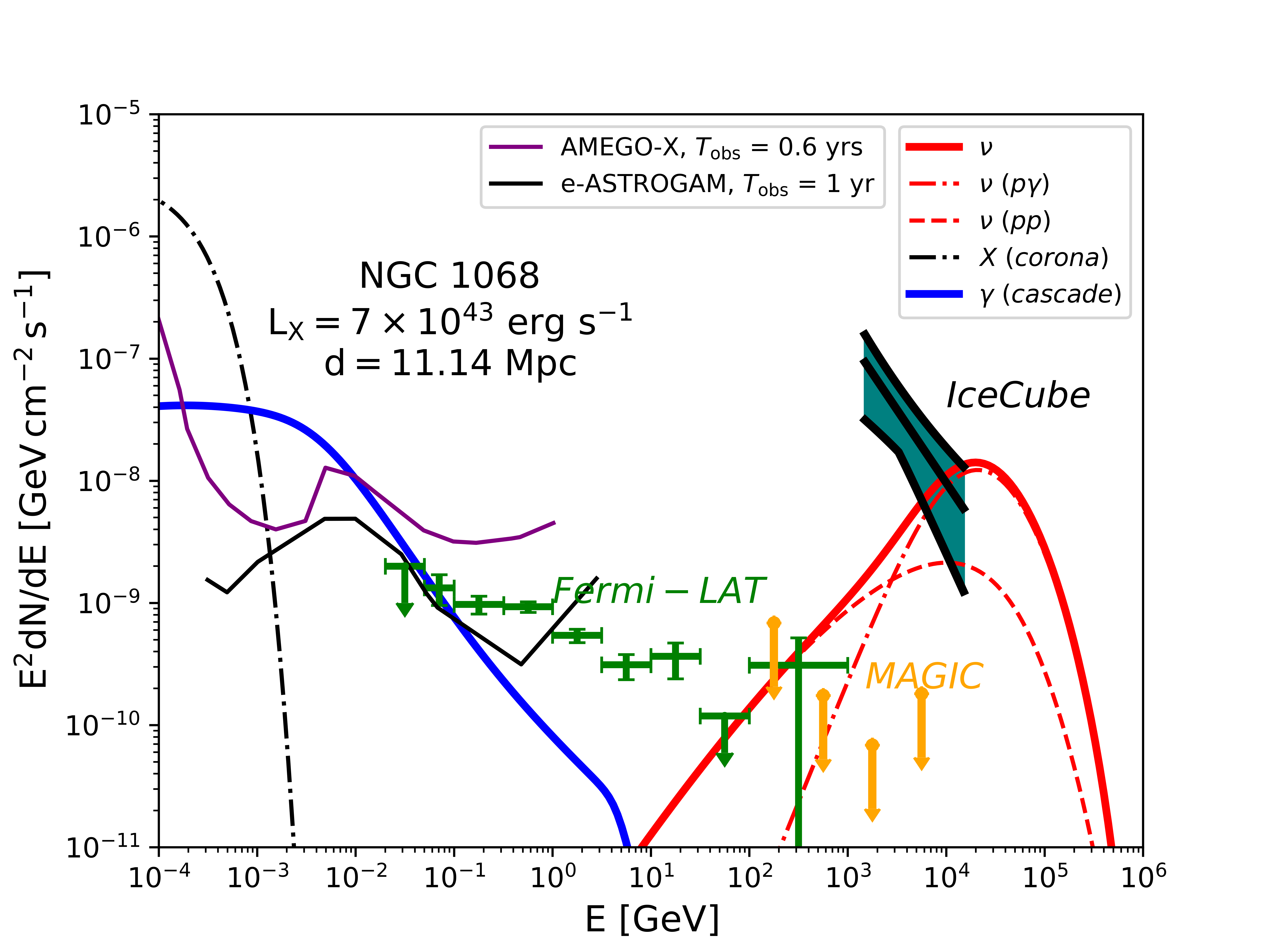}
\caption{The gamma-ray and neutrino spectra emanating from NGC 1068 are presented here.
The black solid lines denote the $95$\% contour lines and the best-fit curve derived from IceCube data \cite{IceCube:2022der}, while the green and orange points represent gamma-ray observations from {\it Fermi}-LAT \cite{Ajello:2023hkh} and MAGIC \cite{MAGIC:2019fvw}, respectively.
The purple and black solid lines represent the sensitivities of AMEGO-X~\cite{Caputo:2022xpx} and e-ASTROGAM~\cite{e-ASTROGAM:2016bph} for observation times of 0.6 years and 1 year, respectively.
The black dash-dotted line illustrates the tail of X-ray emission originating from the corona.
The all-flavor neutrino spectrum (depicted as a red solid line) and the cascade photon spectrum (shown as a blue solid line) are the computational outcomes corresponding to parameters $L_{X} = 7 \times 10^{43}~{\rm erg}~{\rm s}^{-1}$ and $\mathcal{R} = 10$.}
    
\label{fig:NGC1068}
    
\end{center}
\end{figure}
\begin{figure*}[t]
 \includegraphics[width=0.46\textwidth]{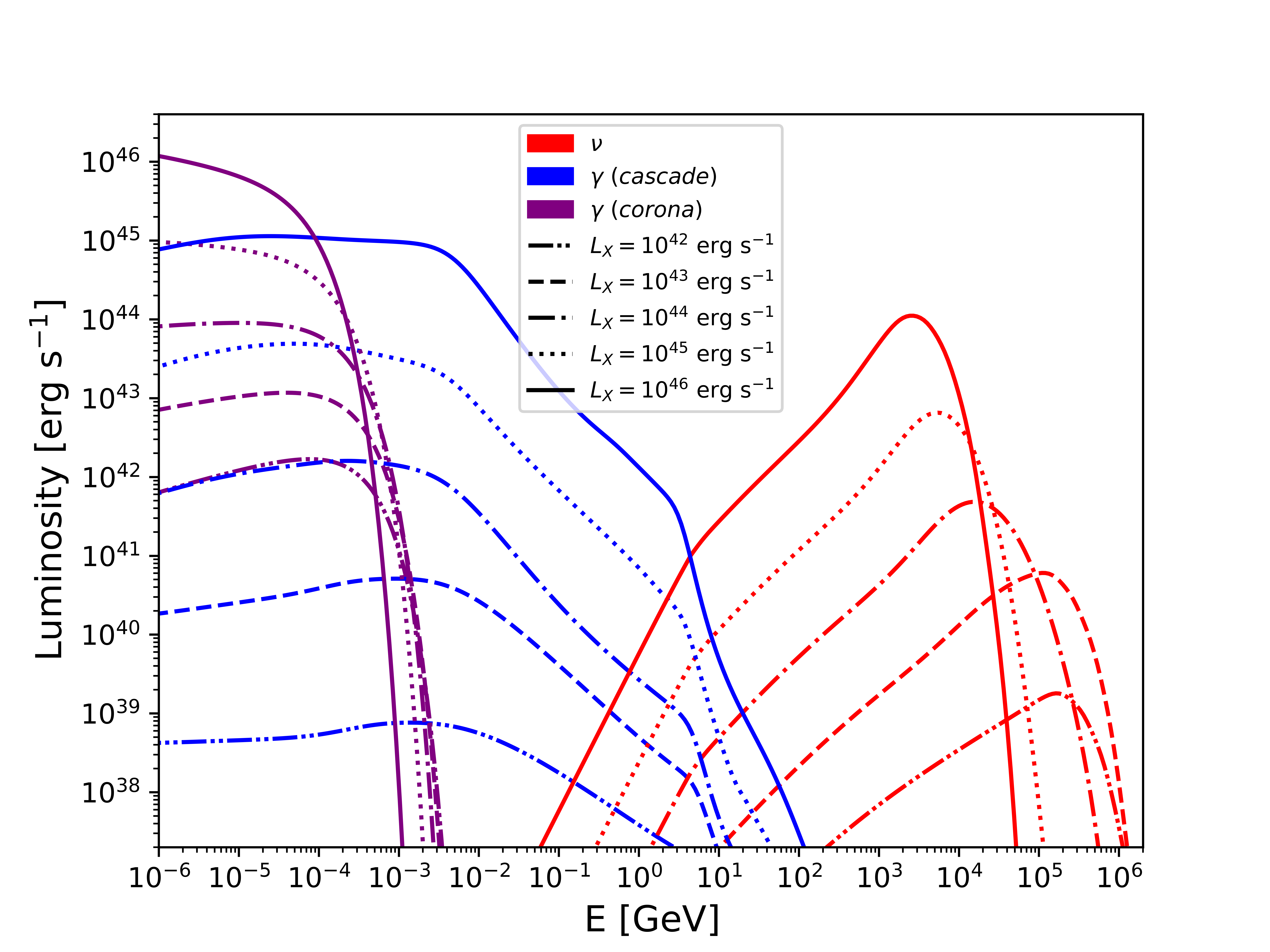}
 \includegraphics[width=0.46\textwidth]{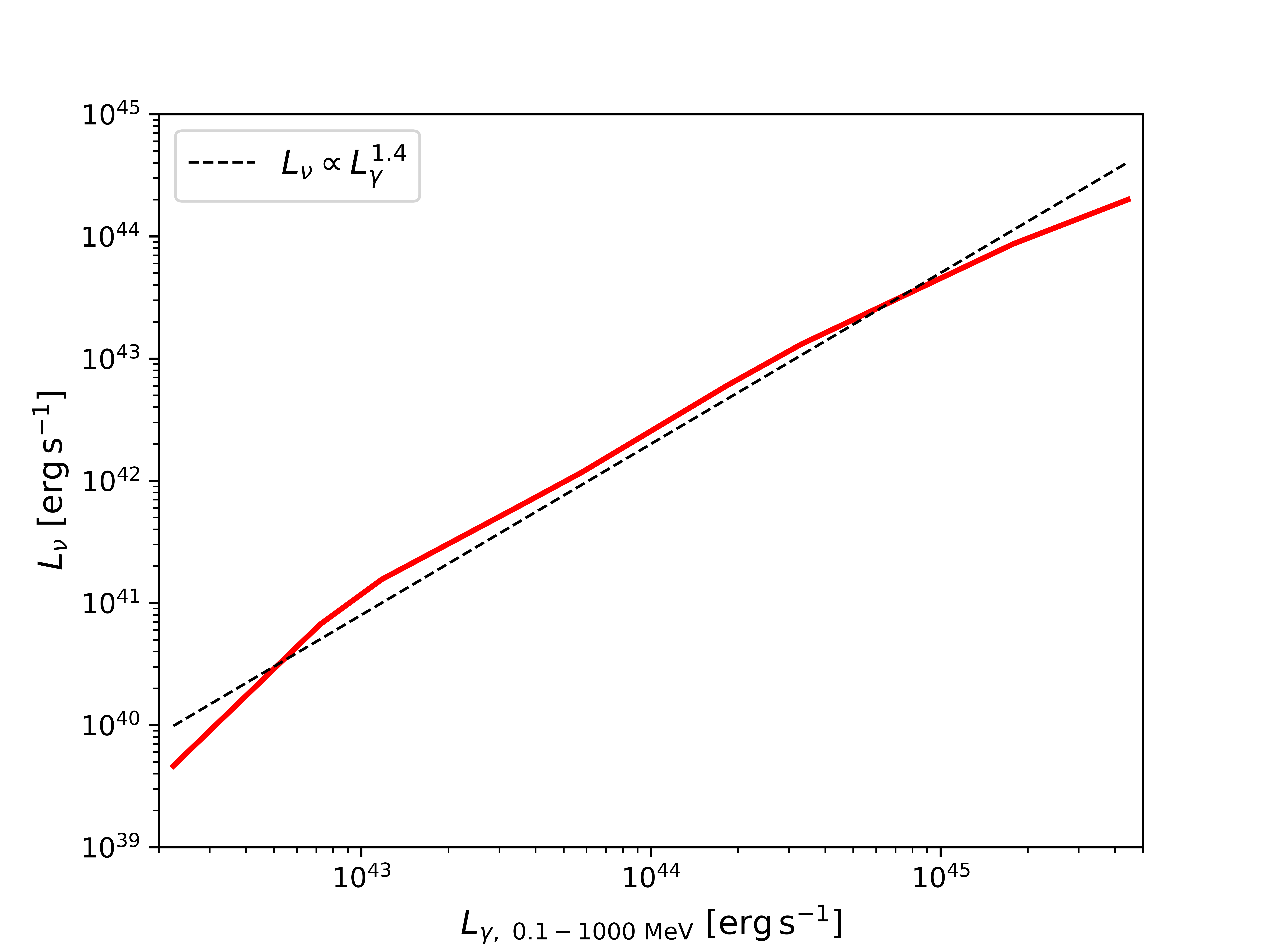}
 \caption{\textit{(Left)} Spectral energy distributions of neutrinos (red) and cascade photons (blue) under different X-ray luminosities correspond to the calculation results with parameter $\mathcal{R}=10$. Curves for different X-ray luminosity values are labeled in the legend. \textit{(Right)} The neutrino luminosity above 300 GeV is shown as a function of the gamma-ray luminosity $L_{\gamma,~{0.1 - 1000}~{\rm MeV}}$, comprising contributions from both coronal and cascade photons.}

\label{fig:spectrum10R}
\end{figure*}

The second term in the Fokker-Planck equation represents continuous energy losses. The total cooling rate for protons is given by the sum of the contributions from all relevant processes $t_{p,\rm cool}^{-1}=t_{pp}^{-1}+t_{p\gamma}^{-1}+t_{\rm BH}^{-1}+t_{p,\rm syn}^{-1}$. The cooling rate due to $pp$ inelastic collisions is estimated by
\begin{equation}
    t^{-1}_{pp} \approx n_p \kappa_{pp} \sigma_{pp} c,
\end{equation}
where $\kappa_{pp}\approx 0.5$ and $\sigma_{pp}$ are the inelasticity and the scattering cross-section for the $pp$ collision process~\cite{Kafexhiu:2014cua, Murase:2013rfa}, respectively. The cooling rate for the photomeson process $(p\gamma)$ and the Bethe-Heitler process $(\rm{BH})$ can be estimated by
\begin{equation}
t_{p\gamma, \rm{BH}}^{-1}\approx\frac{c}{2\gamma_p^2}\int_{\bar{\varepsilon}_{\rm th}}^\infty \hat{\sigma}(\bar{\varepsilon})\hat{\kappa}(\bar{\varepsilon})\bar{\varepsilon}{d}\bar{\varepsilon} \int_{\bar{\varepsilon}/(2\gamma_p)}^\infty \varepsilon^{-2}\frac{dn}{d\varepsilon}{d}\varepsilon,
\end{equation}
where $\gamma_{p}$ is Lorentz factors of protons, $\bar{\varepsilon}$ is the photon energy in the proton rest frame, $dn/d\varepsilon$ is the number density of target photons. The threshold energies of $p\gamma$ and BH process are $\bar{\varepsilon}_{\rm{th}} \approx 0.145~\rm{GeV}$ and $2m_{e}c^{2}$, respectively. $\hat{\sigma}$ and $\hat{\kappa}$ denote cross-section and inelasticity for either the $p\gamma$ process~\cite{Hummer:2010vx, Stecker:1968uc} or the BH process~\cite{Blumenthal:1970nn, 1992ApJ...400..181C}. Finally, the synchrotron radiation cooling of protons is given by
\begin{equation}
    t_{p,\rm syn} \approx \frac{6 \pi m^{4}_{p} c^3}{m^{2}_{e} \sigma_{T} B^2 \varepsilon_{p}}.
\end{equation}

The escape timescale in the escape term is given by~\cite{Fiorillo:2024akm, Fiorillo:2025ehn}
\begin{equation}
    t_{p,\rm esc} \approx \frac{R}{c} \, \mathrm{max} \left[1, \frac{R}{\ell} \left(\frac{e B \ell}{\varepsilon_p}\right)^{1/3} \right].
\end{equation}

For the injection term, We adopt a mono-energetic injection ~\cite{Murase:2019vdl}
\begin{equation}
\dot {\mathcal F}_{p,\rm inj}=\frac{f_{\rm inj}L_X}{4\pi {(\varepsilon_{\rm inj}/c)}^3 {\mathcal V}}\delta(\varepsilon_p-\varepsilon_{\rm inj}),
\end{equation}
where $\varepsilon_{\rm inj}$ is the injection energy, $f_{\rm inj}$ is the injection fraction at $\varepsilon_{\rm inj}$, ${\mathcal V}$ is the volume of the corona, and $\delta(x)$ is the Dirac delta function. The choice of injection spectrum does not affect the resulting spectral shape of the accelerated protons. This is because the steady-state proton spectrum is primarily determined by the balance among stochastic acceleration, energy losses, and escape.

We numerically solve Eq. (\ref{eq:F_p}) using the Chang-Cooper method~\cite{burgess_2018_1163533, CHANG19701, 1996ApJS..103..255P} to obtain the steady-state proton distribution $\mathcal{F}_p$ in momentum space. 
The cooling rates for the $p\gamma$ and BH processes are calculated using the methods implemented in the $\rm{AM}^{3}$ code \cite{Klinger:2023zzv, Gao:2016uld}. 
The $p\gamma$ cooling rate is derived from the code's interaction rate using an inelasticity coefficient $\kappa_{p\gamma} \approx 0.2$~\cite{Murase_2023}.

The resulting steady-state proton spectra are shown in Fig.~\ref{fig:ProtonSpectrum} for the parameter set listed in Table~\ref{tab:quantities}. These spectra reveal a significant softening as $L_X$ increases from $10^{42}~\mathrm{erg}~\mathrm{s}^{-1}$ to $10^{44}~\mathrm{erg}~\mathrm{s}^{-1}$, a trend we attribute to the BH process dominating the proton cooling at higher luminosities. This interpretation is confirmed by the timescale comparison in Fig.~\ref{fig:proton_times}, which shows the BH cooling rate surpassing other rates. The corresponding evolution of key physical quantities for 
$\mathcal{R}=10$ is detailed in Table~\ref{tab:physicalquantities}.

\section{Neutrino and cascade gamma-ray emission from a population of Seyfert galaxies}\label{sec: neutrino and gamma}
\begin{figure*}[t]
 \includegraphics[width=0.47\textwidth]{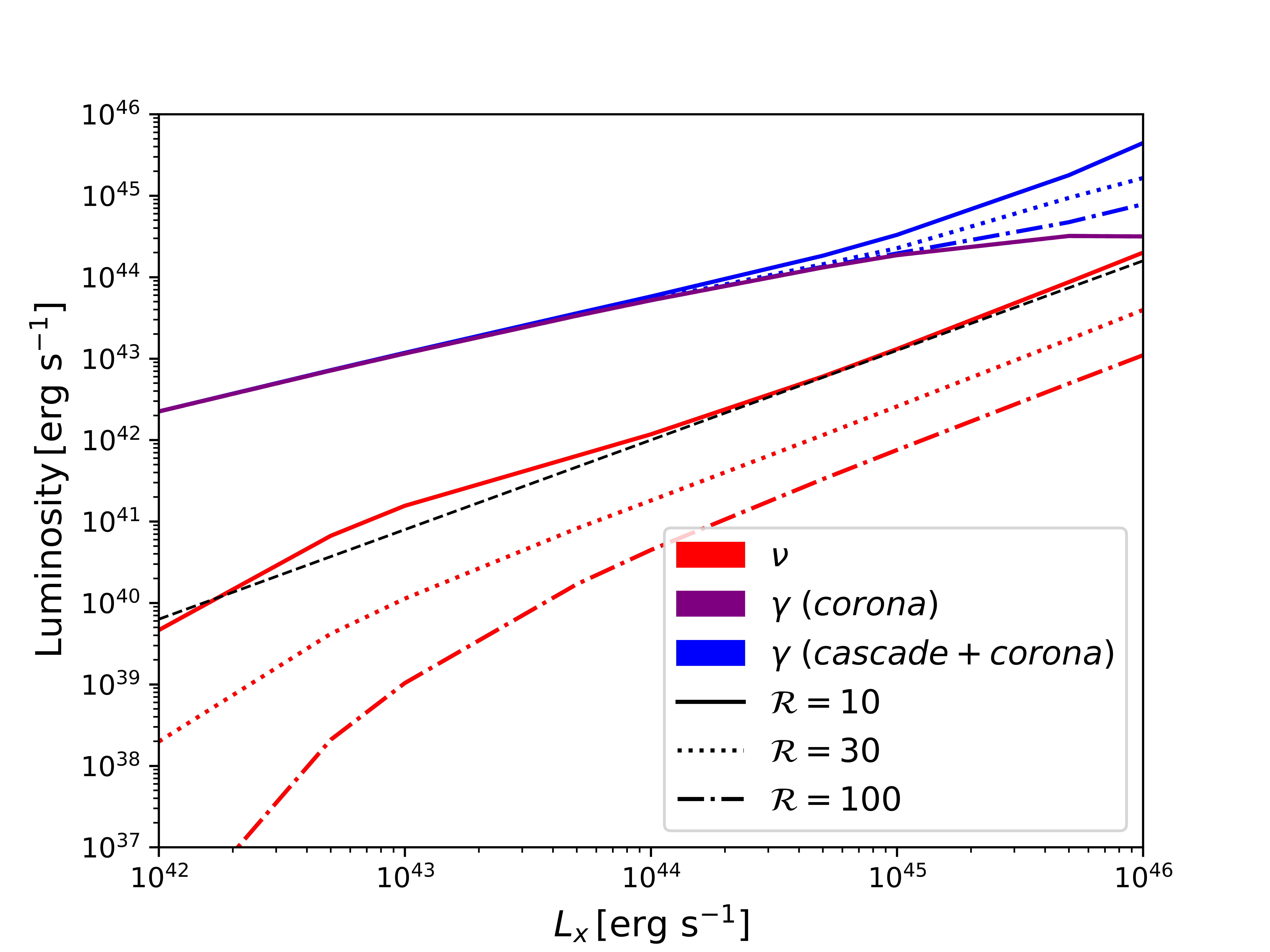}
 \includegraphics[width=0.47\textwidth]{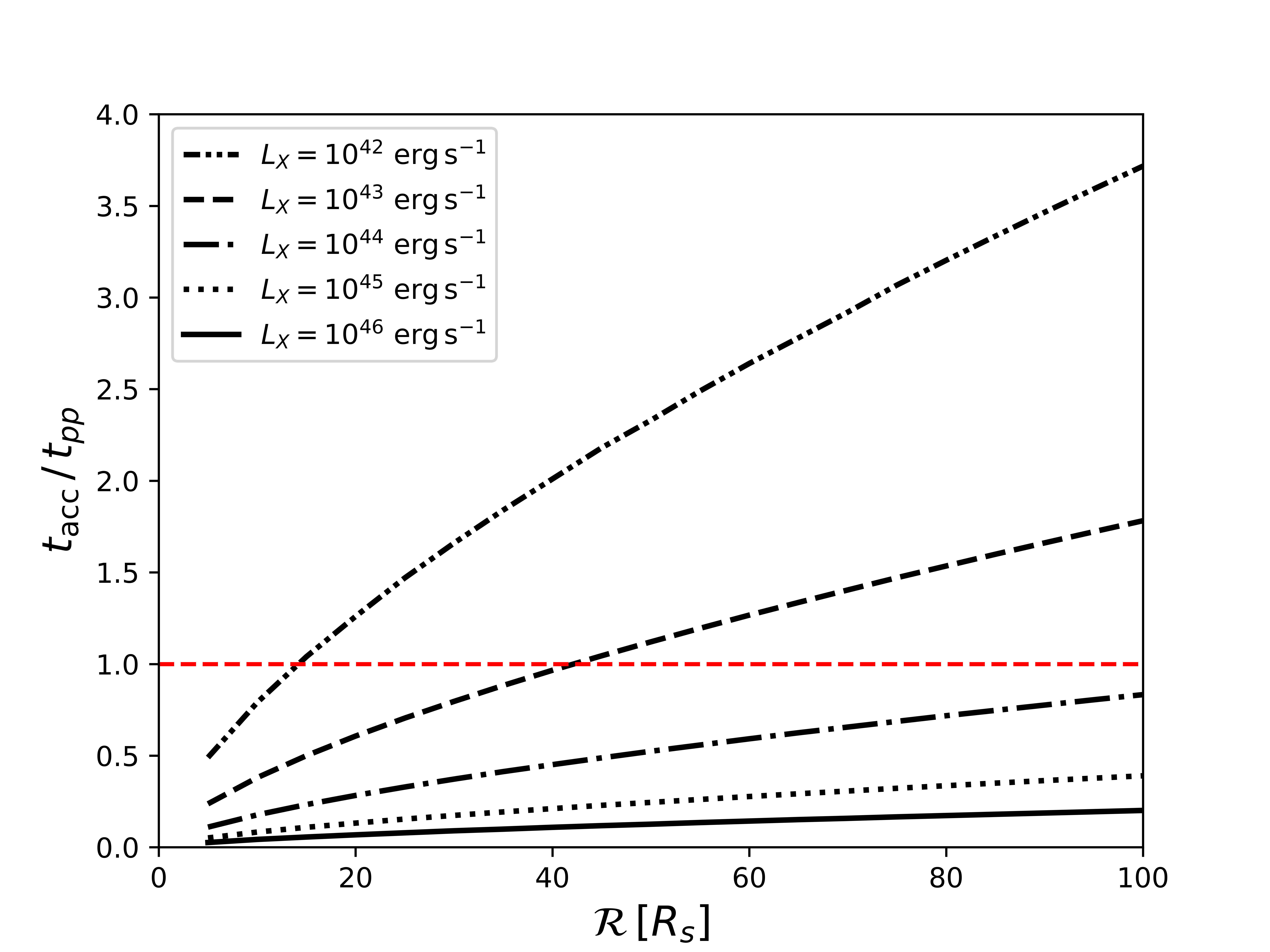}
 \caption{\textit{(Left)} For different values of $\mathcal{R}$, the luminosities of all-flavor neutrinos and gamma rays are shown as a function of the X-ray luminosity $L_{X}$. The red curves represent the neutrino luminosity above 300 GeV, while the blue and purple curves represent the gamma-ray luminosity in the energy range of $0.1-1000$ MeV. The solid black line corresponds to the fit between $L_{\nu}$ and $L_X$ for $\mathcal{R} = 10$, as given in Eq.~(\ref{eq:fit}). Different $\mathcal{R}$ values represented by the line segments are indicated in the legend. \textit{(Right)} For different $L_{X}$, the ratio between the acceleration timescale $t_{\rm{acc}}$ and the cooling timescale $t_{pp}$ is shown as a function of the dimensionless radius $\mathcal{R}$. The legend shows the values of $L_{X}$ for the different curves.}

\label{fig:Luminosity}
\end{figure*}

High-energy protons interact with photons and protons in the corona via processes such as the $p\gamma$ process, the BH process, and the $pp$ process, producing an abundance of secondary particles, including electrons, photons, and neutrinos. Among these, high-energy gamma rays initiate electromagnetic cascades by interacting with the surrounding radiation fields.
Adopting the open-source code $\rm{AM}^3$~\cite{Klinger:2023zzv}, we calculate these processes in the disk-corona model and derive the spectrum of the secondary particles.
Our model is compared with multi-messenger observations~\cite{IceCube:2022der, Ajello:2023hkh, MAGIC:2019fvw} of NGC 1068 in Fig.~\ref{fig:NGC1068}, assuming $L_{X} = 7 \times 10^{43}~{\rm erg}~{\rm s}^{-1}$~\cite{Marinucci:2015fqo} and $\mathcal{R} = 10$ at luminosity distance of 11.14 Mpc~\cite{Tikhonov2021}.
The computed all-flavor neutrino spectrum (Here we assume a neutrino flavor ratio of  $\nu_{e} : \nu_{\mu} : \nu_{\tau} \approx 1 : 1 : 1$ at Earth after oscillations)  exhibits a two-component structure, originating from $pp$ interactions (red dashed line) at lower energies and $p\gamma$ interactions (red dotted-dashed line) at higher energies.
The region of neutrino production is opaque to gamma-rays above the GeV energy band due to two-photon absorption. The optical depth $\tau_{\gamma \gamma}$ can be estimated as~\cite{Murase:2022dog}
\begin{equation}
\begin{aligned}
    \tau_{\gamma \gamma} &\approx \eta_{\gamma \gamma} \sigma_{T} R n_{X} \left(\frac{\varepsilon_{\gamma}}{ \Tilde{\varepsilon}_{\gamma\gamma-e^{+}e^{-}}}\right)^{\Gamma_{X}-1} \\
    & \simeq 194 ~ \left(\frac{L_{X}}{7 \times 10^{43}~{\rm erg}~{\rm s}^{-1}} \right)^{0.42} \left(\frac{\varepsilon_{\gamma}}{1~\rm{GeV}}\right) \mathcal{R}^{-1}_{1},
\end{aligned}
\end{equation}
where $\eta_{\gamma\gamma} \approx 0.12$ is a coefficient~\cite{Svensson:1987nlx}, $\Gamma_{X} \approx 2$, $\Tilde{\varepsilon}_{\gamma\gamma-e^{+}e^{-}} \approx m^{2}_{e}c^4 / \varepsilon_{X} \simeq 0.13~\mathrm{GeV}~(\varepsilon_{X}/2~\mathrm{keV})^{-1}$, and $n_{X}=L_{X}/(4\pi R^2 c \varepsilon_{X})$ is the number density of X-rays.

Fig.~\ref{fig:spectrum10R} (left panel) shows the resulting neutrino and cascade photon spectra as a function of $L_{X}$, derived from the disk-corona photon field and the accelerated proton distribution.
We find that although the neutrino luminosity generally increases with $L_{X}$, its spectral peak shifts to lower energies.
This behavior arises because $L_{X}$ not only supplies more photons for hadronic interactions but also enhances both the $p\gamma$ and Bethe–Heitler cooling processes, thereby suppressing the acceleration of protons to the highest energies.
The primary contribution to cascaded gamma-rays comes from the synchrotron radiation and inverse Compton scattering of electron-positron pairs produced by $\gamma\gamma$ interactions.
As shown in Fig.~\ref{fig:spectrum10R} (left), this MeV cascade component emerges clearly across all $L_{X}$ values, with a luminosity comparable to that of neutrinos. This establishes MeV gamma rays as an indispensable co-probe alongside neutrinos. While the corona is opaque to GeV–TeV photons, MeV emission provides a direct observational signature of the hidden hadronic processes. Consequently, future telescopes such as AMEGO-X~\cite{Caputo:2022xpx} and e-ASTROGAM~\cite{e-ASTROGAM:2016bph} will be able to critically test this model by searching for the predicted MeV counterpart from sources like NGC 1068 (Fig. \ref{fig:NGC1068}), providing essential observational constraints on the connection between neutrinos and cascade photons.

In Fig.~\ref{fig:spectrum10R} (right panel), we also show the relationship between the neutrino luminosity above 300 GeV and the gamma-ray luminosity in the energy range of 0.1-1000 MeV, where the gamma-ray emission includes contributions from both coronal and cascade components. 
The integral energy range of 0.1-1000 MeV is chosen based on the observational capabilities of future MeV telescopes such as AMEGO-X~\cite{Caputo:2022xpx}.
The relationship between the neutrino luminosity and the gamma-ray luminosity is not linear. For $\mathcal{R}=10$, we find a scaling
\begin{equation}\label{eq:fit}
    L_\nu\simeq 2\times 10^{42}~{\rm erg}~{\rm s^{-1}}~\left(\frac{L_{\gamma,~{0.1 - 1000}~{\rm MeV}}}{10^{44}~{\rm erg}~{\rm s^{-1}}}\right)^{1.4}.    
\end{equation}
This $L_\nu-L_{\gamma,\rm 0.1-1000 MeV}$ relation provides a practical observational tool. Since both neutrinos and MeV gamma rays are co-produced in the hadronic interaction chain, this scaling provides a unified multi-messenger signature of the underlying process, which will be directly testable with future MeV telescopes.

In Fig.~\ref{fig:Luminosity} (left panel), we present the dependence of the luminosities of neutrinos ($E_{\nu}>300 {\rm GeV}$) and gamma rays ($0.1-1000$ MeV) on the X-ray luminosity $L_{X}$ for different corona size $\mathcal{R}$. For a fiducial size of $\mathcal{R} = 10$, the relation between $L_{\nu}$ and $L_X$ is given by 
\begin{equation}\label{eq:fit}
    L_\nu\simeq 1\times 10^{42}~{\rm erg}~{\rm s^{-1}}~\left(\frac{L_{X}}{10^{44}~{\rm erg}~{\rm s^{-1}}}\right)^{1.1}.    
\end{equation}
This empirical relation provides a direct link between the X-ray luminosity and the predicted neutrino flux. 
It allows astronomers to estimate the expected neutrino flux from a Seyfert galaxy based on its X-ray luminosity. This predictive capability is crucial for guiding targeted searches with current and future neutrino telescopes like KM3NeT~\cite{KM3NeT:2022pnv}, IceCube~\cite{IceCube:2013llx} and IceCube-Gen2~\cite{IceCube-Gen2:2020qha}, significantly enhancing the signal-to-noise ratio in likelihood-based analyses by providing a prior weighting for potential neutrino sources.
Future measurements in the X-ray, MeV gamma-ray, and neutrino bands will powerfully constrain the coronal model parameters, such as the magnetic field strength, turbulence level, and most importantly, the corona size $\mathcal{R}$, thereby discriminating between different acceleration and emission scenarios.

Furthermore, our model also indicates that the size of the corona should be compact, since at larger $\mathcal{R}$, the $pp$ cooling process suppresses the proton acceleration. 
Fig.~\ref{fig:Luminosity} (right) illustrates this by plotting the $t_{\rm acc}/t_{pp}$ ratio on $\mathcal{R}$ for various values of $L_{X}$.
These results, as established in our analysis, depend on the model parameters $\eta$, $\sigma_{\rm{tur}}$ and $\tau_{T}$.

\section{Summary}\label{sec: summary}
Motivated by the recent IceCube observations, we have constructed a self-consistent model for high-energy neutrino and gamma-ray production in the magnetized, turbulent coronae of Seyfert galaxies.
We adopt the non-resonant acceleration mechanism recently identified by PIC simulations, which is fundamentally distinct from traditional resonant acceleration. 
This mechanism features an energy-independent acceleration rate governed by the largest turbulent structures, enabling more efficient proton acceleration. 
Our framework synthesizes the standard disk-corona spectral energy distribution with key empirical relations, resulting in a model that is primarily parameterized by the X-ray luminosity $L_{X}$ and the dimensionless normalized radius $\mathcal{R}$. 

By numerically solving the Fokker-Planck equation for proton transport and simulating subsequent hadronic interactions, we find that in regimes of high $L_{X}$ and compact size (small $\mathcal{R}$), the increased target photon density can lead to the production of cascade photon luminosity comparable to that of the target photon.
In Seyfert galaxies, since proton acceleration and cooling occur within the same region, we infer that a compact corona is required to ensure that the acceleration process can effectively counteract the cooling processes. This conclusion aligns with previous findings from model-independent multi-messenger constraint studies~\cite{Murase:2022dog}. Although such a compact corona effectively attenuates high-energy gamma-rays, our calculation confirms that MeV gamma rays can efficiently escape from the corona. This establishes MeV gamma rays and neutrinos as complementary probes of the hidden non-thermal processes, in which neutrinos directly trace proton interactions, while MeV emission traces the reprocessed electromagnetic cascade.

We further analyze the dependence of neutrino and cascade photon luminosities on $L_{\rm X}$ and $\mathcal{R}$. For corona with size of $\mathcal{R}=10$, the neutrino luminosity exhibits a nearly-linear relationship with X-ray luminosity, fitted as $L_{\nu} \propto L_X^{1.1}$, which can be used to weight potential neutrino sources based on their observed X-ray flux. Additionally, a correlation is found between neutrino luminosity and gamma-ray luminosity in the 0.1–1000 MeV band, described by $L_{\nu} \propto L_{\gamma, 0.1-1000{\rm MeV}}^{1.4}$ for $\mathcal{R}=10$. This offers a testable prediction for next-generation MeV telescopes such as AMEGO-X~\cite{Caputo:2022xpx} and e-ASTROGAM~\cite{e-ASTROGAM:2016bph}.

The spectral and luminosity characteristics of neutrinos in our study are governed by the competition between acceleration and cooling. The BH process plays a particularly critical role by setting the maximum proton energy and thereby directly shaping the neutrino spectrum, highlighting the model's sensitivity to the target photon field. Its competition with turbulent acceleration explains the non-linear luminosity scaling at low $L_{\rm X}$, while at larger coronal radii, 
$pp$ cooling becomes the primary suppression mechanism. 

Future measurements in the X-ray, MeV gamma-ray, and neutrino bands, greatly enhanced by the capabilities of next-generation instruments like AMEGO-X~\cite{Caputo:2022xpx}, e-ASTROGAM~\cite{e-ASTROGAM:2016bph}, and IceCube-Gen2~\cite{IceCube-Gen2:2020qha}, will powerfully constrain the coronal model parameters, such as the magnetic field strength, turbulence level, and the corona size.
At the same time, refining the disk-corona radiation environment will be essential to achieve precise, source-specific predictions. These efforts will jointly discriminate between different acceleration and emission scenarios.

\section{ACKNOWLEDGEMENTS}

H.N.He is supported by Project for Young Scientists in
Basic Research of Chinese Academy of Sciences (No. YSBR-061), and by NSFC under the grants No. 12173091, No. 12333006, and No.12321003, and the Strategic Priority Research Program of the Chinese Academy of Sciences No.XDB0550400. 
C.Q.Pang is supported by the National Natural Science Foundation of China under Grants No.~11965016.

\bibliographystyle{apsrev4-1}
\bibliography{main}

\end{document}